\documentstyle[10pt,epsf]{article}

\include{epsf}
\epsfverbosetrue
\newcommand{\half}
             {\mbox{\small $\frac{1}{2}$}}  

\begin{document}

\message{Kyffhaeuser Talk, \today}

\thispagestyle{empty}

\vspace{-1cm}
\begin{flushright}
   HLRZ 93-69  \\
   October 28, 1993
\end{flushright}

\begin{center}
   \begin{Large}
      {\bf{The $\pi N$ Sigma term -- a lattice investigation}}%
      \footnote{The $MT_c$ Collaboration.}
      \footnote{Talk presented by R. Horsley at the
                Workshop on Field-Theoretical Aspects of Particle Physics,
                Kyffh\"auser, Germany,
                September $13^{th}$ to $17^{th}$ $1993$. 
                To be published in the NTZ Series.}
   \end{Large} \\[0.50cm]
    R. Altmeyer$^{a}$, M. G\"ockeler$^{b}$, R. Horsley$^{b,a}$, 
    E. Laermann$^{c}$ and G. Schierholz$^{a,b}$                    \\[0.25cm]
    {\small $^a$DESY, Notkestra{\ss}e 85,
                D-22603 Hamburg, Germany}   \\
    {\small $^b$HLRZ, c/o Forschungszentrum J\"ulich,
                D-52425 J\"ulich, Germany}     \\
    {\small $^c$Fakult\"at f\"ur Physik, Universit\"at Bielefeld,
                D-33501 Bielefeld, Germany} 
\end{center}

\vspace{0.25cm}

\centerline{\bf Abstract}
\noindent
A lattice calculation of the $\pi N$ sigma term is described using dynamical
staggered fermions. Preliminary results give a sea term comparable in
magnitude to the valence term.

\vspace{0.20cm}
\noindent
{\bf Introduction and Theoretical Discussion}

Hadrons appear to be far more complicated than the (rather successful)
constituent quark model would suggest. For example a recent result
from the EMC experiment, \cite{ashman89a}, suggests that constituent
quarks are responsible for very little if any of the nucleon spin.
Another, older result, is from the $\pi N$ sigma term, see e.g.
\cite{hoehler83a}, which seems to give a large strange component
to the nucleon mass. To explain theoretically these
results and other experiments involves the computation of certain
matrix elements -- a non-perturbative calculation. This is thus an area where
lattice calculations may be of some help.

The $\pi N$ sigma term, $\sigma_{\pi N}$, is defined%
\footnote{Actually, originally as the matrix element of the double
commutator of the Hamiltonian with two axial charges.
However this is equivalent to the definition given
in eq.~(\ref{theory.a}), see for example ref.~\cite{cheng88a}.}
as that part of the mass of the nucleon (for definiteness the proton)
coming from the vacuum connected expectation value of the up ($u$)
and down ($d$) quark mass terms in the QCD Hamiltonian,
\begin{equation}
   \sigma_{\pi N} = m \langle N | \bar{u}u+\bar{d}d | N \rangle ,
\label{theory.a}
\end{equation}
where we have taken these quarks to have equal current mass ($=m$).
Other contributions to the nucleon mass come from the chromo-electric
and chromo-magnetic gluon pieces and the sea terms due to the $s$ quarks.
Experimentally this matrix element has been measured from low energy
$\pi$-$N$ scattering. A delicate extrapolation to the chiral limit
\cite{cheng88a,cheng71a} gives a result for the isospin even
amplitude of $\Sigma/f_\pi^2$ with $\Sigma = \sigma_{\pi N}$,
from which the $\pi N$ sigma term may be found. The precise value
obtained this way has been under discussion for many years.
However within the limits of our lattice calculation, this will
not concern us here and for orientation we shall just quote a range
of results from later analyses of
$\sigma_{\pi N} \approx 56 \mbox{MeV}$, \cite{gasser88a}, down to
$45 \mbox{MeV}$, \cite{gasser91a}.

To estimate valence and sea contributions to $\sigma_{\pi N}$,
classical current algebra analyses assume octet dominance and make
first order perturbation theory about the $SU_F(3)$ flavour symmetric
Hamiltonian. We have
\begin{equation}
   \hat{\cal H}_m = m(\bar{u}u + \bar{d}d) + m_s\bar{s}s 
                  \equiv
                   m_{av} \bar{\psi}I \psi
                   - {{m_s-m}\over{\sqrt{3}}} \bar{\psi}\lambda^8 \psi   .
\label{theory.b}
\end{equation}
The first term in the second equation is the flavour
symmetric Hamiltonian with mass $m_{av} = (2m+m_s)/\sqrt{3}$, while the
second term is the flavour breaking piece. ($\psi$ is the column
vector $(u,d,s)$.) We thus have from eq.~(\ref{theory.a})
\begin{eqnarray}
   \sigma_{\pi N}
              &\approx& m \langle N | \bar{u}u+\bar{d}d -2\bar{s}s
                                    |N \rangle_{symm}
                        +  2m\langle N |\bar{s}s |N \rangle_{symm}
                                                           \nonumber \\
              &\stackrel{def}{=}& \sigma^{val}_{\pi N} + \sigma^{sea}_{\pi N},
\label{theory.c}
\end{eqnarray}
where we have first assumed that the nucleon wavefunction does not change
much around the symmetric point. We then subtract and add a strange
component. At the symmetric point the $u$ and $d$ quarks each have equal
valence and sea part, while the $s$ quark matrix element only has a sea
component. Thus in the first term the sea contribution cancels, justifying
the definitions given in the last line of eq.~(\ref{theory.c}). Also
this means that in the following we shall interchangeably talk about 
`strange' and `sea' contributions to the nucleon. At first order 
perturbation theory $\sigma^{val}_{\pi N}$ may easily be calculated%
\footnote{Baryon octet mass splittings are given by
\begin{displaymath}
   M_B - M_{symm} = 
        - (m_s-m)\langle B|\bar{\psi}\lambda^8\psi|B\rangle_{symm}/\sqrt{3} ,
\end{displaymath}
where $B$ is the $\half^+$ nucleon multiplet. From the Wigner-Eckart
theorem for $SU(3)$ we have
\begin{displaymath}
   \langle B |\bar{\psi}\lambda^8\psi|B\rangle_{symm}
       = F tr (B^\dagger[\lambda^8,B]) + D tr(B^\dagger\{\lambda^8,B\}) .
\end{displaymath}
From the known baryon masses $F$, $D$ can then be estimated.
(We use the numbers given in \cite{gupta91a}
as $(m_s-m)F \approx 190$, $(m_s-m)D \approx -61$.) Furthermore
\begin{displaymath}
   \sigma_{\pi N}^{val} = m\sqrt{3} \langle N|\bar{\psi} \lambda^8\psi
                                    | N \rangle_{symm}
                                 = m(3F - D) .
\end{displaymath}
Using $m_s/m \approx 26$ gives the quoted results.}
to give $\sigma^{val}_{\pi N} \approx 25 \mbox{MeV}$ and so
$\sigma^{sea}_{\pi N} \approx 31 \sim 20 \mbox{MeV}$.
This in turn means that
\begin{equation}
   m_s \langle N | \bar{s}s | N \rangle \approx 400 \sim 250 \mbox{MeV},
\label{theory.d}
\end{equation}
which would indicate a sizeable portion of the nucleon mass comes from
the strange quark contribution. (Remember that the mass of the proton
is about $938\mbox{MeV}$.) This would not be expected from the constituent
quark model, where the nucleon is made only of (dressed) $u$ and $d$ 
quarks.

\vspace{0.20cm}
\noindent
{\bf Staggered Fermions}

As we see from above, we need to calculate certain matrix elements.
One of the most promising methods at present to calculate them
comes from the lattice, where Euclidean space-time is discretised
with a lattice spacing $a$. The problem is then turned into a
statistical mechanical one of evaluation of correlation functions
(as to be described below). This can be attacked using Monte Carlo
simulations. The continuum limit is obtained when $a \to 0$,
or $\beta \equiv 6/g^2(a) \to \infty$. This procedure is well known,
see for example \cite{creutz88a}. We have performed a large scale
simulation on a $16^3\times 24$ lattice of QCD using the
Hybrid Monte Carlo method. The main emphasis was for $\beta = 5.35$,
$m=0.01$, although some other runs were also performed.
(Our results and further details are given in \cite{altmeyer93a}.)

There are two commonly used formulations of fermions on the lattice
-- Wilson and staggered. Both have their advantages and disadvantages.
We have used the staggered approach, which describes in the continuum
$4$ degenerate flavours. Here the fermion fields in the partition function,
$\chi^a(x)$ are only a function of the lattice point and colour index.
Quark spin and flavour combinations are given by summing appropriately
over $\chi$'s sitting on the vertices of hypercubes, thus
\begin{equation}
   q_{\alpha f}(X) = \sum_A \Gamma^A_{\alpha f} \chi(X+A).
\label{staggered.a}
\end{equation}
The hypercube is defined by $x= X +A$, $A_\mu$ being $0$ or $1$ and
$X_\mu$ even, where $x$ is the lattice site. $\Gamma^A$ is given
by $\gamma_1^{A_1} \ldots \gamma_4^{A_4}$. Eq.~(\ref{staggered.a})
projects out the flavour ($f$) and spin ($\alpha$) degrees of freedom.
In the continuum limit we have for $m = 0$ a chiral symmetry
$SU_L(4)\otimes SU_R(4)$. The diagonal symmetry, valid for all $m$, is the
flavour group $SU_F(4)$. The advantage of the staggered lattice
fermion formulation is that we have, for $m=0$, a remnant
$U(1)\otimes U(1)$ symmetry.
This allows a clean study of chiral symmetry breaking as $m \to 0$,
and also the emergence of a massless $\pi$, with $m_\pi^2 \propto m$.
(For Wilson fermions, one must carefully tune the hopping parameter to
obtain a massless $\pi$, chiral properties always being explicitly broken.)
As we saw in the last section we have at the symmetric point an $SU_F(3)$
symmetry for the $u$, $d$, $s$ quarks. (Corrections for the strange
quarks are treated in first order perturbation theory, as part of the
sea contribution.) As we shall be here interested in the questions of
sigma/nucleon and valence/sea ratios, we expect that the different number
of quarks present will make little difference.

Lattice calculations have to be performed when the lattice spacing $a$
is small enough that finite lattice effects are negligible. Staggered
fermions have a natural indicator to show this by when the flavour symmetry
is restored. On the lattice meson ($M \sim \bar{\chi}\chi$) and baryon 
($B \sim \chi\chi\chi$) operators which in the continuum describe
degenerate states lie in several distinct lattice representations,
each with a different mass. As $a \to 0$ these masses must tend to
a common mass. Practically we say that we are close enough to the
continuum when possible mass differences are negligible. For our
configurations we seem to be entering such a region, \cite{altmeyer93a}.

Another technical point concerns the existence of a parity partner
in the correlation functions. Masses ($M_\alpha$) and amplitudes
($A_{\alpha\alpha}$) are measured from $2$-point correlation functions.
For the baryon we have
\begin{eqnarray}
   C(t) &\stackrel{def}{=}&
               \langle B(t) \bar{B}_W(0) \rangle
                                                \nonumber \\
        &\approx& \langle 0| \hat{B} \hat{S}_4^t \hat{\bar{B}}_W | 0\rangle
                                                \nonumber \\
        &\approx& A_{NN} \mu_N^t + A_{\Lambda\Lambda} \mu_\Lambda^t
                        \qquad\qquad \half T \gg t \gg 0 .
\label{staggered.b}
\end{eqnarray}
In the second line we have gone from the (numerically) calculable
$2$-point correlation function to the equivalent expression in the
operator formalism. The transfer matrix is denoted by
$\hat{T} = \hat{S}_4^2 \equiv \exp{(-2\hat{H})}$. A shift of one unit
still lies in the hypercube, which then involves a flavour transformation.
Thus we can write $\hat{S}_4 = \hat{\Xi}_4 \hat{T}^{\half}$ where
$\hat{\Xi_4}$ is a non-local flavour transformation. (No explicit 
representation of $\hat{\Xi}_4$ is at present known.) Eigenstates 
of $\hat{S}_4$ are defined by
\begin{equation}
   \hat{S}_4 |\alpha\rangle = \mu_\alpha |\alpha\rangle \qquad 
                              \mu_\alpha = \xi_\alpha e^{-M_\alpha} ,
\label{staggered.c}
\end{equation}
($\langle\alpha|\alpha\rangle = 1$) where $\xi_\alpha = \pm1$ is the
eigenvalue of $\hat{\Xi}_4$. Inserting complete sets of states in the
second line of eq.~(\ref{staggered.b}) gives the third line
where the amplitude
$A_{\alpha\beta} = \langle 0|\hat{B}|\alpha\rangle
\langle\beta|\hat{\bar{B}}_W|0\rangle$, and
$\mu_N = \exp{(-M_N)}$, $\mu_\Lambda = - \exp{(-M_\Lambda)}$.
(We shall call $\Lambda$, a little artificially, the parity
partner to the nucleon.) The time box length ($T$, here $=24$) has been
assumed to be so large that there are no finite time-size affects
(although these can be easily taken into account).
The parity partner arises because on the lattice inversion
$I_s : x_i \to -x_i$ is not parity ($P$) as it effects the flavour
degrees of freedom. (One can show that $\hat{P} = \hat{I}_s \hat{\Xi}_4$.)
As a consequence hadron operators in general do not have a definite parity.

\begin{figure}[tbh]
\vspace*{-1.5cm}
\hspace*{3.0cm}
\epsfxsize=12.5cm \epsfbox{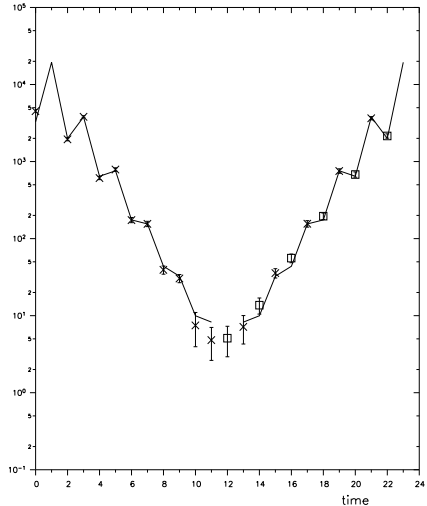}
\vspace*{-1.0cm}
\caption{A typical baryon correlation function plotted against
         $t$ for $\beta=5.35$, $m=0.01$. Boxes denote negative values.
         The fit interval is from $t=2$ to $22$.}
\label{figbar}
\end{figure}

As an example of a correlation function we show in Fig.~\ref{figbar}
a baryon correlation function. The typical zigzagging of the results
(and fit) is caused by the parity partner. To pick out the ground states
we must take $t \to \infty$ and/or choose an operator with a good
overlap with this state. Experience has shown that a non-local operator
(the `wall' and hence the subscript `$W$') has rather good overlap
properties with the ground state for very moderate $t$, so this has
been used as the source; the sink is taken as a local baryon operator.
The mass obtained this way agrees with that which comes from using
a local baryon operator for both sink and source.

\vspace{0.20cm}
\noindent
{\bf Measuring $\sigma_{\pi N}$}

\begin{figure}[htb]
\vspace*{-1.5cm}
\hspace*{3.0cm}
\epsfxsize=12.5cm \epsfbox{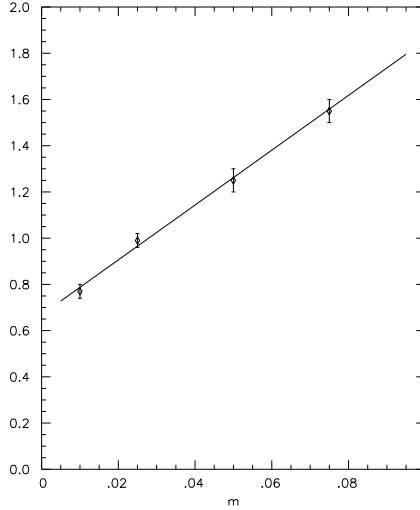}
\vspace*{-1.0cm}
\caption{The Nucleon mass, $M_N$, plotted against $m$ for $\beta=5.35$.}
\label{figgrad}
\end{figure}

Practically there are several possibilities open to us for the evaluation
of the matrix element. The easiest is simply to differentiate
eq.~(\ref{staggered.c}). As%
\footnote{In the staggered formalism the operator expression for
$m\bar{\chi}\chi$ has a more complicated appearance than that given here,
\cite{altmeyer93a,smit91a}.}
${{\partial \hat{S}_4} \over {\partial m}} = - \bar{\chi}\chi \hat{S}_4$
we have
\begin{equation}
   m {{\partial M_N}\over {\partial m}}|_\beta 
              = \langle N|m\bar{\chi}\chi|N\rangle = \sigma_{\pi N},
\label{measure.a}
\end{equation}
an application of the Feynman-Hellmann theorem. Thus all we need to
do is to measure $M_N$ for different masses $m$ (at constant $\beta$)
and numerically estimate the gradient. In Fig.~\ref{figgrad}
we show such a procedure. (At present we have simply fitted a straight
line through all points.) We find
\begin{equation}
   \sigma_{\pi N} \approx 11.9(8)m|_{m=0.01} \approx 0.12(1),
\label{measure.b}
\end{equation}
which gives
\begin{equation}
   {\sigma_{\pi N} \over M_N} \approx
                     {{11.9m}\over {11.9m+0.64}}|_{m=0.01} \approx 0.15,
\label{measure.c}
\end{equation}
which is to be compared with the experimental result of 
$\sigma_{\pi N} / M_N \approx 0.06 \sim 0.05$. The numerical result
is much larger, but presumably this simply indicates that we have
used much too large a quark mass in our simulation (which at present
we, and everybody else, cannot avoid doing due to computer limitations).

We now wish to estimate the valence and sea contributions. This is
technically more complicated and involves the evaluation of a $3$-point 
function, \cite{maiani87a}
\begin{equation}
   C(t;\tau) = \langle B(t) m\bar{\chi}\chi(\tau) \bar{B}_W(0) \rangle ,
\label{measure.d}
\end{equation}
This may be diagrammatically sketched as the sum of two terms:
\begin{figure}[h]
\hspace*{6.0cm}
\epsfxsize=4.0cm \epsfbox{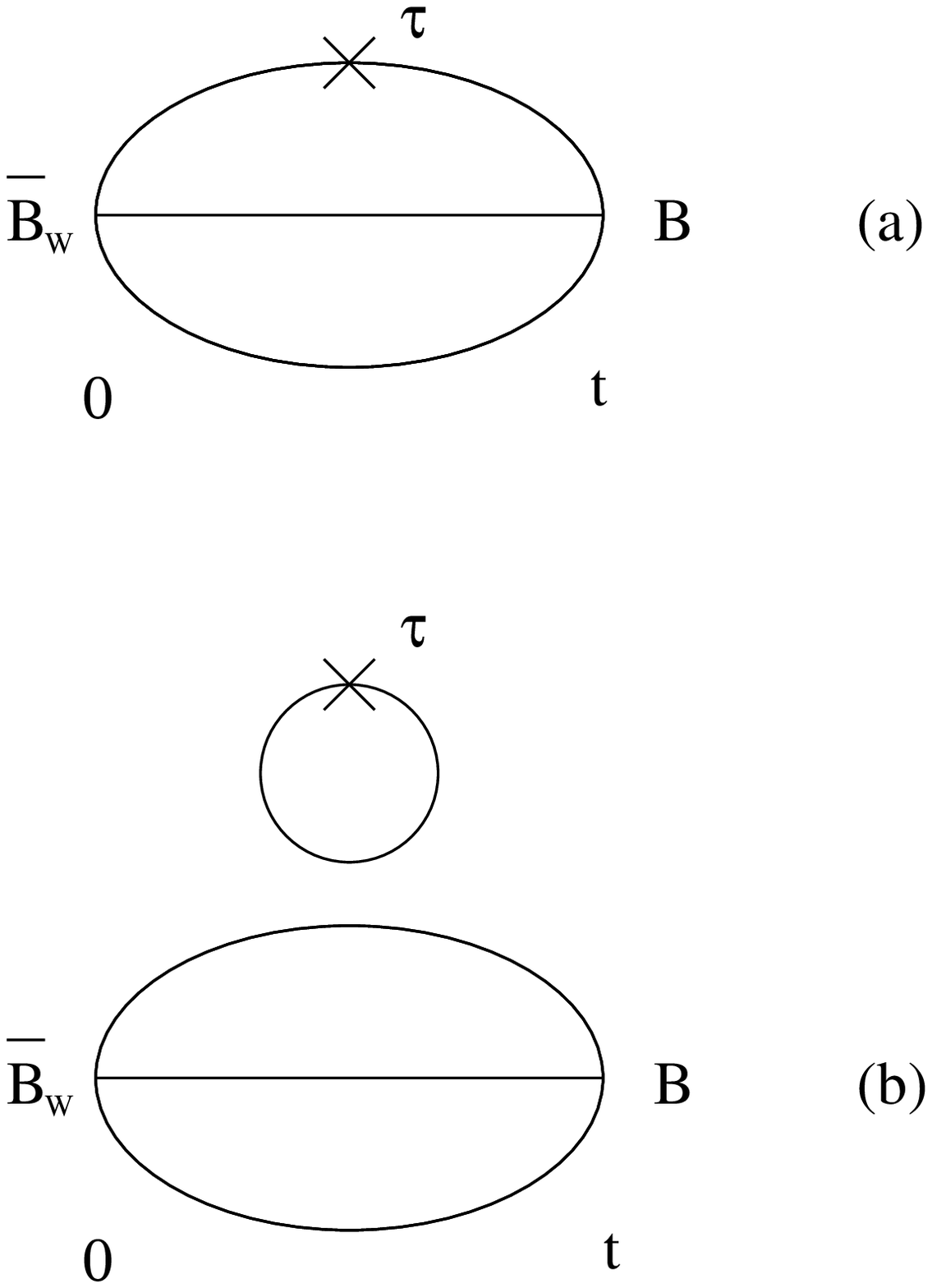}
\label{figsketch}
\end{figure}

\noindent
Diagram (a) is the connected (or valence) piece; the nucleon quark 
lines are connected with the mass insertion operator. Diagram (b)
represents the disconnected (or sea) piece.

\begin{figure}[htb]
\vspace*{-1.5cm}
\hspace*{3.0cm}
\epsfxsize=12.5cm \epsfbox{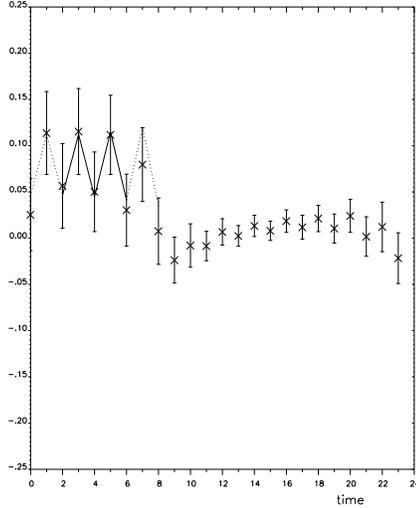}
\vspace*{-1.0cm}
\caption{The ratio of the 3-point connected correlation function to the
         2-point correlation function against $\tau$ for $\beta=5.35$,
         $m=0.01$. The points used in the fit are joined together.
         (The dotted lines just extend the fit to neighbouring points.)}
\label{figconn}
\end{figure}

For the connected piece we have fixed $t$ to be $8$ or $9$
and then evaluated $C(t;\tau)$ as a function of $\tau$. The appropriate
fit function is
\begin{equation}
   C(t;\tau) \approx \sum_{\alpha, \beta = N,\Lambda} A_{\alpha\beta}
                       \langle \alpha| m\bar{\chi}\chi |\beta\rangle
                       \mu_\alpha^{t-\tau} \mu_\beta^\tau
   \qquad\qquad {\half T} \gg t \gg \tau \gg 0.
\label{measure.e}
\end{equation}
The $A_{\alpha\beta}$ and $M_\alpha$ are known from $2$-point  correlation
functions. We see that when $\alpha =\beta =N$ we have the matrix element
that we require: $\langle N|m\bar{\chi}\chi|N\rangle$. However
there are other terms which complicate the fit:
$\langle \Lambda|m\bar{\chi}\chi|\Lambda\rangle$ and the cross terms
$\langle \Lambda|m\bar{\chi}\chi|N\rangle \equiv
\mu_\Lambda\mu_N^{-1}\langle N|m\bar{\chi}\chi|\Lambda\rangle$.
As can be seen from eq.~(\ref{measure.e}) these cross terms are
responsible for oscillations in the result. To disentangle the 
wanted result from $\langle \Lambda|m\bar{\chi}\chi|\Lambda\rangle$
we need to make a joint fit to the $t=8$ and $t=9$ results. At present
we have not done this, but just checked that separate fits give 
consistent results. In Fig.~\ref{figconn} we show preliminary
results for $t=8$. Up until now we have only evaluated about a quarter
of our available configurations, so the final results should have
reduced error bars. We find
\begin{equation}
   \sigma_{\pi N}^{val} \approx 0.08(2),
\label{measure.f}
\end{equation}
with about the same result for the $\Lambda$ matrix element.
The cross term is smaller, roughly $0.02$.

\begin{figure}[htb]
\vspace*{-1.5cm}
\hspace*{3.0cm}
\epsfxsize=12.5cm \epsfbox{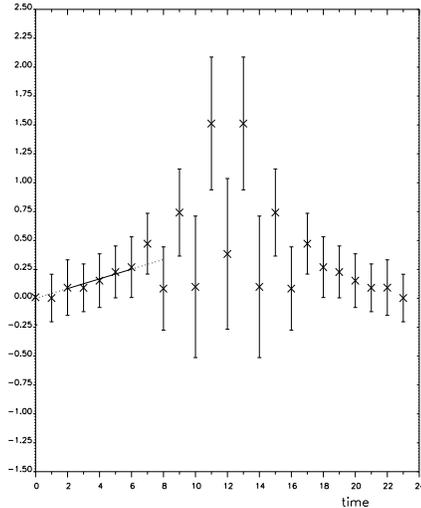}
\vspace*{-1.0cm}
\caption{The ratio of the $3$-point disconnected correlation function
         to the $2$-point correlation function against $t$.
         $100$ sets of Gaussian random numbers were used for the
         stochastic estimator. To improve the signal we have averaged
         over the points $t$ and $T-t$.}
\label{figdisc}
\end{figure}

Finally we have attempted to estimate the disconnected term
using a stochastic estimator, \cite{bitar89a}.
This is costly in CPU time. One can improve the statistics
by summing the 3-point correlation function over $\tau$; this is then
equivalent to a differentiation of the 2-point function with respect
to $m$. This gives
\begin{equation}
   \sum_\tau C(t;\tau)
      \approx t \sum_{\alpha=N,\Lambda}
                    A_{\alpha\alpha}
                    \langle\alpha| m\bar{\chi}\chi
                    |\alpha\rangle  \mu_\alpha^t
       \qquad\qquad \half T \gg t \gg 0 .
\label{measure.g}
\end{equation}
In Fig.~\ref{figdisc} we show the disconnected part of the 3-point
correlation function. As expected the quality of the data is poor --
optimistically we see a slope. A simple linear fit gives
\begin{equation}
   \sigma_{\pi N}^{sea} \approx 0.05(4) .
\label{measure.h}
\end{equation}

\vspace{0.20cm}
\noindent
{\bf Discussion}

We see that (roughly at least) eq.~(\ref{measure.b}) is consistent
with eqs.~(\ref{measure.f},\ref{measure.h}) and that
\begin{equation}
   { \sigma_{\pi N} \over \sigma_{\pi N}^{val} } \approx 1.5 ,
\label{discussion.a}
\end{equation}
which is to be compared with the experimental value of about
$2.2 \sim 1.8$. Although we can draw no firm conclusions at present
our result tentatively indicates that the valence part of the
$\pi N$ sigma term is slightly larger than the sea term.

Comparing our results with other work obtained using dynamical
fermions, Gupta et. al, \cite{gupta91a} who use $2$ flavours of
Wilson fermions, find
$\sigma_{\pi N} / \sigma_{\pi N}^{val} \approx 2 \sim 3$, which
indicates a somewhat larger sea component in the $\pi N$ sigma term.
On the other hand Patel, \cite{patel92a}, using results from \cite{brown91a}
for $2$ staggered flavours finds for the ratio $1.5 \sim 2.0$,
while Bernard et. al., \cite{bernard93a}, have $\sim 2.0$.
These, like our result, seems to be lower than for the Wilson fermion case.

We would also like to emphasise that although lattices can give a
first principle calculation, at present one is not able to do this.
Technically the fit formul{\ae} that we employ, eqs.~(\ref{measure.e},
\ref{measure.g}) are true only for the complete correlation function.
However we have used them for either the connected or the disconnected
part separately. The best way to circumvent this problem is to make 
simulations at different strange quark masses; differentiation as in
eq.~(\ref{measure.a}) would then give directly
$m_s\langle N|\bar{s}s|N\rangle$, the strange content of the nucleon.
However this calculation is not feasible at the present time.
This should be the ultimate goal of lattice simulations,
as other recent (non-lattice) theoretical results,
\cite{gasser91a}, have hinted that perhaps the strange quark content of the
nucleon is not as large as supposed, previous results being explained
by a combination of factors, such as $\Sigma \ne \sigma_{\pi N}$
and higher order corrections to first order perturbation theory.
(Indeed there are already tantalising lattice indications that this may
be so, \cite{guesken88a,sommer91a}.)

In conclusion we would just like to say that lattice results at present
are generally in qualitative agreement with other theoretical and
experimental results. However much improvement in the calculations
is required to be able to make quantitative predictions.

\vspace{0.20cm}
\noindent
{\bf Acknowledgements}

This work was supported in part by the Deutsche
Forschungsgemeinschaft. The numerical computations were performed
on the Cray Y-MP in J\"ulich with time granted by the Scientific
Council of the HLRZ. We wish to thank both institutions for their support.

\end{document}